\documentclass[aps,prl,reprint,groupedaddress,nofootinbib]{revtex4-1}
\usepackage{hyperref}
\usepackage{amsmath}
 \usepackage{multirow}
\usepackage{array}
\newcolumntype{L}[1]{>{\raggedright\let\newline\\\arraybacksslash\hspace{0pt}}m{#1}}
\newcolumntype{C}[1]{>{\centering\let\newline\\\arraybackslash\hspace{0pt}}m{#1}}
\newcolumntype{R}[1]{>{\raggedleft\let\newline\\\arraybackslash\hspace{0pt}}m{#1}}
\usepackage{float}
\usepackage{graphicx,epstopdf}
\usepackage{epsfig}
\usepackage{subfigure} 
\usepackage{psfrag}
\usepackage{color} 
\usepackage{slashed}

\usepackage{amsfonts} 
\usepackage{bbm}
\usepackage{bm}
\usepackage{amssymb}
\usepackage{tikz}
\usepackage{tikz}
\usetikzlibrary{positioning,arrows}
\usetikzlibrary{decorations.pathmorphing} 
\usetikzlibrary{decorations.markings}

\newcommand*{\be}{\begin{equation}}
\newcommand*{\ee}{\end{equation}}

\newcommand{\comment}[1]{}

\newcommand{\cref}[1]{Chapter~\ref{c.#1}}

\def\beq{\begin{eqnarray}}
\def\eeq{\end{eqnarray}}
\def\ba{\begin{array}}
\def\ea{\end{array}}
\def\bi{\begin{itemize}}
\def\ei{\end{itemize}}
\def\be{\begin{enumerate}}
\def\ee{\end{enumerate}}
\def\bc{\begin{center}}
\def\ec{\end{center}}
\def\bt{\begin{table}}
\def\et{\end{table}}
\def\btb{\begin{tabular}}
\def\etb{\end{tabular}}

\def\lsim{\raise0.3ex\hbox{$\;<$\kern-0.75em\raise-1.1ex\hbox{$\sim\;$}}}
\def\gsim{\raise0.3ex\hbox{$\;>$\kern-0.75em\raise-1.1ex\hbox{$\sim\;$}}}

\usepackage[normalem]{ulem}

\begin{document}

\title{Massive Gravitons as Feebly Interacting Dark Matter Candidates}

\author{Haiying Cai$^{1}$}
\email{hcai@korea.ac.kr  (corresponding author)}
\author{Giacomo Cacciapaglia$^{2,3}$}
\email{g.cacciapaglia@ipnl.in2p3.fr}
\author{Seung J. Lee$^{1}$}
\email{sjjlee@korea.ac.kr}

\affiliation{
$^1$Department of Physics, Korea University, Seoul 136-713, Korea\\
$^2$University of Lyon, Universit\'e Claude Bernard Lyon 1,
F-69001 Lyon, France\\
$^3$Institut de Physique des 2 Infinis de Lyon (IP2I), UMR5822, CNRS/IN2P3,  F-69622 Villeurbanne Cedex, France}

\begin{abstract}
We detailed our discovery of  a chiral enhancement in the production cross sections of massive spin-2 gravitons, below the electroweak symmetry breaking scale, that makes them ideal dark matter candidates for the freeze-in mechanism. The result is independent of the physics at high scales, and points toward masses in the keV- MeV range. The graviton is, therefore, a sub-MeV dark matter particle, as favored by the small scale galaxy structures. We apply the novel calculation to a Randall-Sundrum model with multiple  branes, showing a significant parameter space where the first two massive gravitons saturate the dark matter relic density.
\end{abstract}

\maketitle

Despite the overwhelming evidence for the presence of a dark matter (DM) component in our Universe, also indirectly observed in galaxies and galaxy clusters, the nature of this matter component remains a mystery. In the Standard Model (SM), no known particle can play the role of DM: the only candidates, neutrinos, have a relic density many orders of magnitude below the required one, which is roughly 5 times the relic density of ordinary baryons.
Extended objects, like primordial black holes, remain a possibility, alas still requiring new physics to explain their presence~\cite{Cheong:2019vzl}.

A particle DM candidate can only emerge from new physics beyond the SM. The most popular and time honored possibility has been the  weakly interactive massive particle (WIMP), which requires substantial interactions with the SM particles, with an annihilation cross section in the order of the electroweak ones, $\sigma v \simeq 2.0 \times 10^{-26} cm^3/s$, independent of the DM mass in order to achieve the observed  relic density $\Omega_{\rm DM} h^2 \sim 0.12$~\cite{Aghanim:2018eyx}.  The non-observation of new physics signals at colliders (the LHC) and at DM direct and indirect detection experiments  has, however, put this scenario under stress \cite{Bertone:2018krk}. Hence, this `WIMP crisis', has prompted the exploration of alternative possibilities. Here we will be interested in the freeze-in mechanism of feebly interacting massive particles (FIMPs), which never attain thermal equilibrium with the SM bath~\cite{Hall:2009bx}.  This is usually obtained by tuning a coupling to very small values, typically of the order of $10^{-8}-10^{-10}$ provided the DM is  stable (protected by a parity). Another possibility, called ultraviolet (UV) freeze-in~\cite{Elahi:2014fsa}, relies on higher dimensional operators suppressed by a large scale. In the latter case, the naive expectation is that unitarity violation renders the DM relic density highly sensitive to the unknown reheat temperature at the end of the inflationary phase. Here, we will present a novel scenario based on a spin-2 DM candidate where, albeit the freeze-in is induced by a non-renormalizable operator, the UV sensitivity of the predictions is curbed by the electroweak scale. The mechanism behind this is a chiral enhancement of some scattering amplitudes involving massive SM fermions, which is only activated below the electroweak symmetry breaking (EWSB) scale, $T_{\rm EW} \approx 160$~GeV.

The couplings of the spin-2 state (a.k.a. massive graviton) $G_{\mu \nu}$ to the SM particles can be parametrized by the following 4-dimensional effective Lagrangian:
\beq
\mathcal{L}_{eff} = \sum_{i = \mbox{spin}} C_i \, G^{\mu \nu}  \left(2 \frac{\delta \mathcal{L}^i}{\delta \hat g^{\mu \nu}} -  \eta_{\mu \nu} \mathcal{L}^i \right)|_{\hat g = \eta}\,, \label{Lag0}
\eeq
where $\eta$ is the Minkowski metric, and the factors within parentheses are the stress-energy tensors $T_{\mu \nu}^i$ for the SM particles of different spins ($i = 0, \frac{1}{2}, 1$). As such, $\mathcal{L}^i$ is the SM Lagrangian for the particles of spin $i$ and, since  the spin sum  of the graviton polarization $P_{\mu \nu, \alpha \beta} =  \frac{1}{2} \left( P_{\mu \alpha} P_{\nu \beta}  +   P_{\nu \alpha} P_{\mu \beta}   -  \frac{2}{3}P_{\mu \nu} P_{\alpha \beta} \right) $, with $P_{\mu \nu } = \eta_{\mu \nu } -\frac{k_\mu k_\nu}{M_G^2}$  is traceless, the terms proportional to $\mathcal{L}^i$ in Eq.~\eqref{Lag0} do not contribute.  Explicit expressions for the various spins can be found in Ref.~\cite{Han:1998sg}. Note that the massive spin-2 DM candidate is not the mediator of gravitational interactions, which are generated by the usual massless gravitons in our scenario. Universal couplings between the massive graviton $G_{\mu \nu}$ and all the SM particles ensures unitarity in the high energy limit before the chiral symmetry breaking. Hence, we will stick to this assumption and denote the universal coupling as $C_H$.
The results we present here are independent on the origin of the degrees of freedom needed to give mass to the spin-2 state, while later we will focus on a class of 5-dimensional models, where the mass is generated by the compactification of the extra space.

The freeze-in  generally proceeds  via  decays of heavier particles or pair annihilations of  the SM particles in the thermal bath. The FIMP belongs to a hidden sector that communicates to the SM sector via a superweak portal. Assuming that the inflaton dominantly decays into the SM particles, the initial DM abundance after reheating can be neglected, so that  the FIMP  relic density is produced  via an accumulation process. In our simplified model,  the massive graviton is a perfect FIMP candidate for DM due to the smallness of the gravitational coupling $C_H$. As we will see, however, as the massive graviton can decay via the same coupling, $C_H$ is forced to be too small to produce a sizable density of gravitons. Here we will point out a mechanism in place below the EWSB scale that invalidates this conclusion. First, we introduce the Boltzmann Equation (BE)  describing the production of the FIMP graviton in the early Universe. For a generic scattering process  $B_1 + B_2 \to B_3 + G$, where $B_{1,2,3}$ are the SM particles, the evolution of the massive graviton number density $n_G$ follows the BE~\cite{Hall:2009bx}:  
\beq
&& \frac{d n_G}{d t}+3 H n_G   \approx  \frac{  T}{512 \pi^6}\int_{s_0}^{\infty} ds \, P_{B_1B_2}P_{B_3 G} \nonumber \\   & & \quad  \times \mathcal{A}_{B_1B_2\rightarrow B_3 G} K_1(\sqrt{s}/T)/ \sqrt{s} \,; \label{BE}
\eeq
where  $H$ is the Hubble expansion parameter defined as the time derivative of the logarithmic scale factor $\ln a(t)$, $K_1$ is the first modified Bessel function of the second kind and
\beq
& & P_{B_i B_j} = \frac{(s-(m_{B_i}+m_{B_j})^2)^{\frac{1}{2}} (s-(m_{B_i}-m_{B_j})^2)^{\frac{1}{2}}}{2 \sqrt{s}}\,,  \nonumber \\  
& & \mathcal{A}_{B_1B_2\rightarrow B_3 G} =  \int d \Omega |\mathcal{M}|^2_{B_1B_2\rightarrow B_3 G}\,,  \nonumber \\
& & s_0 = \mbox{max} ~\{(m_{B_1}+m_{B_2})^2, (m_{B_3}+ m_G)^2\}\,.
\eeq
with $\mathcal{A}_{B_1B_2\rightarrow B_3 G}$  standing for the amplitude squared after the solid angle integration and $s = (p_{B_1} + p_{B_2})^2$.  The right-hand side  of Eq.~(\ref{BE}) is the interaction rate density $\gamma (T) $ for DM production where the thermal average is performed by the technique developed in~\cite{Gondolo:1990dk}. 
By solving the BE with the proper initial condition, one finds that the FIMP relic density is directly proportional to the $2\to 2$ cross section, in contrast to the inverse proportionality in freeze-out models. At high temperatures, above the EWSB scale, we find that all amplitudes squared scale like $ \mathcal{A}_{B_1B_2\rightarrow B_3 G^{(1)}}  \sim C_H^2 \; g_{i}^2\; s$, where $g_i$ is an appropriate SM coupling. The only exception is the process $h h \to h G$, for which the amplitude is a constant  $ \mathcal{A}_{h h \to h G} \propto C_H^2 \frac{m_{h}^4}{v^2}$.  As $C_H \sim M_{\rm Pl}^{-1}$ is  suppressed by the Planck mass to ensure Hubble timescale stability, the cross-sections are too small to provide a  DM-like relic density via $2\to2$  scattering freeze-in.

In this Letter we discovered a chiral enhancement of a class of processes that is active below the EWSB scale, and in the limit of light graviton. Thus, in the following we will assume that $M_G$ is much smaller than any other scale in the process. The chirally enhanced processes involve SM fermions and a massless gauge boson. The most dominant ones, therefore, involve quarks and gluons: $q \bar{q} \to g G$, $q g \to q G$, etc. Note that the process with heavy lepton pairs and one photon will also undergo this chiral enhancement, but suppressed by a factor of $e^2/(4 g_{s}^2)$ in the amplitude squared. For the process $\bar q q \to g G$, the total amplitude squared  before EWSB  is  calculated (see the Supplementary material) using the Feynman Rules in~\cite{Giudice:1998ck, Han:1998sg}:
\beq
\mathcal{A}^0_{\bar q q   } = \frac{128 \pi}{3}  C_H^2 g_s^2 s  \,, \label{Aqq0}
\eeq
where $g_s$ is the chromodynamic coupling and the result is consistent with the results in~\cite{Mirabelli:1998rt} in the limit of $M_G \to 0$.  In contrast, after EWSB, the leading term in the small $M_G$ expansion is given by:
\beq
\mathcal{A}_{\bar q q  } &=&  \frac{256 \pi C_H^2 g_{s}^2 m_{q}^2 s \left(s + 2 m_{q}^2\right)}{9 M_{G}^4} \,. \label{Aqq}
\eeq
This term comes from the contribution of the longitudinal polarization of the massive graviton in terms of $\frac{k_{\mu} k_{\nu}}{M_G^2}$, which contributes only after a chirality flip of the fermion line via a mass insertion. The enhancement is more effective for heavy quarks (the bottom and charm, as the top is too heavy to be in thermal equilibrium below the EWSB) and can overcome the Planck suppression for $M_G$ in the keV--MeV range.
For the process $q  g \to q  G$, we can use the cross symmetry to get the amplitude squared:
\beq
\mathcal{A}_{q g  } = \mathcal{A}_{\bar q g  } &=& \frac{256 \pi C_H^2 g_{s}^2 m_{q}^2 \left(s- m_{q}^2\right)^2 
   \left(s+ m_{q}^2\right)}{3 s M_G^4 } \,. \label{Aqg}
\eeq
Note that, when inserting the amplitudes squared in the BE \eqref{BE}, we need to consider the renormalization group evolution of the coupling constant,  $\alpha_s (\mu) = \frac{g_s^2}{4 \pi}$ where $\mu = \sqrt{s}$. In the numerical results, we will use the following one-loop running coupling
\beq
\frac{1}{\alpha_s(\mu)} = 8.47 + \frac{7}{2 \pi} \log\left(\frac{\mu}{ M_Z} \right)\,.
\eeq

We are now ready to compute the massive graviton relic density.
After reheating,  the Universe entered into the radiation-dominated era, with the reheating temperature $T_{\rm RH}$ representing the maximum temperature reached by the thermal bath~\cite{Cheung:2011mg, Hannestad:2004px}. The freeze-in production of the massive graviton can be divided into two phases: the UV phase above the EWSB scale and the infrared (IR) one after EWSB, i.e. for temperatures above and below the critical temperature $T_{C} \simeq 160$ GeV~\cite{Quiros:1999jp}.  The SM particles are  in thermal equilibrium  after reheating, while the massive graviton is  not. This is because the superweak portal ensures $ \gamma(T) <  H  n_\gamma^{eq} $ both in the UV and IR phases.  As a simple estimation, after EWSB,  for the massive graviton to stay in non-equilibrium requires  $T^3 \lesssim  \pi^6 M_{\rm Pl}  \frac{M_G^4}{m_q^2}$.  Provided the FIMP mass is  $M_G \sim 1$ MeV, this translates to an upper bound  $T_C \lesssim 1$ TeV.  Rewriting the BE in terms of the yield $Y_G = n_G / S$, with $S$ being the entropy density,  the IR contribution to the freeze-in density can be written as:
\beq
&& Y_{\rm IR}  \simeq  \frac{1}{2048 \pi^6} \int_{T_{ QCD}}^{T_{C}} \frac{d T}{S H} \bigg(  \int_{4 m_q^2}^{\infty} d s  (s - 4 m_q^2)^{1/2} A_{\bar q q } \nonumber \\ && K_1\left(\frac{\sqrt{s}}{T} \right)     + 2 \int_{m_q^2}^{\infty} d s   \frac{(s -  m_q^2)^2}{ s^{3/2}} A_{q g }   K_1\left(\frac{\sqrt{s}}{T}\right) \bigg)\,, \label{YIR}
\eeq
with
\beq
S = \frac{2 \pi^2 g_{*}^s T^3 }{45} \,, \quad  H = \sqrt{\frac{g_{*}^{\rho} \pi^2 }{90}}  \frac{T^2}{M_{pl}}                    
\eeq
where $M_{pl}$ is  the reduced Planck mass for the Hubble parameter in the radiation-dominated era, $T_{QCD} \simeq 150$ MeV and  $g_{*}^s \simeq g_{*}^\rho \simeq 10^2 $.  As typically  $T_{RH} > T_C$, the IR contribution is not sensitive to the reheating temperature $T_{RH}$. Moreover, the temperature integration is dominated by the interval close to  $T_C$ with the IR yield  proportional to $m_q^2$. Hence,  even though the processes involving lepton pairs can continue to  temperatures below the QCD phase transition,  their contribution remains subleading.  By evaluating Eq.(\ref{YIR}) numerically and taking into account the  bottom and charm quarks,  we obtain the following  result:
\beq
\Omega_{\rm IR } h^2 &= & \frac{M_{G}}{3.6 \times 10^{-9} ~\mbox{GeV}} Y_{\rm IR} \nonumber \\ 
&\simeq&  3.0 \times 10^{31} ~\mbox{GeV}^5 ~\frac{C_H^2}{M_{G}^3}  \,. \label{OmegaIR}
\eeq
Since the couplings of  $G$  to the SM particles are  model independent and uniquely dictated by  symmetries, we can ignore the self-interactions after including a radion-like field $r$. The decay width of a graviton of a few MeV mass is governed by  the Lagrangian in  Eq.~\eqref{Lag0}:
\beq
 && \Gamma (G \rightarrow e^+ e^- +  \nu_i \bar{\nu}_i  +  \gamma\gamma)  \simeq     \frac{9 C_{H}^2 M_G^3}{320 \pi} \, \label{tot}
 \eeq
Now we can  estimate the upper bound of the IR freeze-in contribution by combining Eq.~(\ref{OmegaIR}) with the lifetime  $\tau_G = 6.58 \times 10^{-25}/\Gamma_G$  from  Eq.~(\ref{tot}),
\beq
\Omega_{\rm IR} h^2 \lesssim 0.12 \times \left( \frac{1.6~ \mbox{MeV}}{M_G} \right)^6 \frac{10^{27}~\mbox{Sec}}{\tau_G} \,. \label{bound}
\eeq
Given  the  larger branching ratio of  graviton into $\gamma \gamma$ than into $e^+ e^-$,  we impose an appropriate lifetime limit $\tau_G \gtrsim 10^{27}$ sec, deriving from the stringent bounds from indirect detection and cosmic microwave background (CMB)~\cite{Essig:2013goa, Poulin:2016anj}. Hence, in Eq.(\ref{bound}), for each fixed value of $M_G$,  the maximum relic density is given by the largest value of $C_H$ that satisfies the lifetime bound. We found, therefore, that the DM relic density can be saturated by a single massive graviton with $M_G \lesssim 1.6$~MeV, with increasing lifetimes (decreasing $C_H$) for smaller masses.

The UV contribution can be computed with a similar formula to Eq.~\eqref{YIR}, taking into account all the quarks in the SM (which are massless in this regime). Numerically, we obtain
\beq
\Omega_{\rm UV } h^2 &= &  1.2 \times 10^{29} ~\mbox{GeV} ~C_H^2 ~ M_{G}  \nonumber \\
&\simeq &   \Omega_{\rm IR} h^2  \left( \frac{ M_G}{4.0 ~\mbox{GeV}} \right)^4 \,. \label{OmegaUV}
\eeq
Note that we used $T_{RH} = 10^6$ GeV as a template value \cite{Cheung:2011mg} in the above estimate. Due to the linear divergence in $s$ of  Eq.~\eqref{Aqq0}, the  integration of BE gives  $\Omega_{\rm UV}h^2 \propto T_{RH}$.  One can anticipate,  for  $M_G \sim \mathcal{O} (1)$  MeV, that the UV result is at most $10^{-14}$  times the IR one  and much smaller in case of  a lower  $T_{RH}$ input.


\begin{figure}[tb]
	\centering  
	\includegraphics[height=6.0 cm,  
		width=7.2 cm]{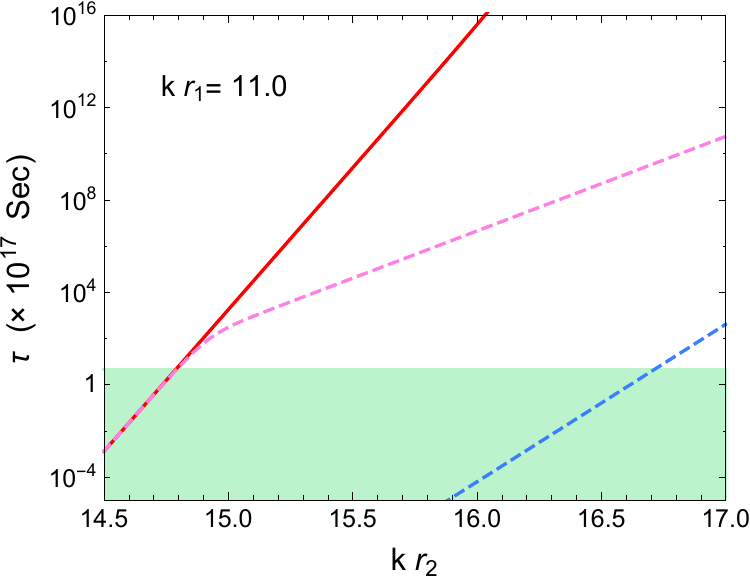}  
		\caption{The  lifetime for the lightest KK graviton (red lines) and the IR brane radion (blue line) as a function of $k r_2$, with  $k r_1 = 11.0$.  The solid line corresponds to $m_r > 2 m_G^{(1)}$ whereas the dashed line is for  $m_r = 3. 45 \Lambda_{\rm IR}$. The region with lifetime below the age of  the Universe ($\sim H_0^{-1}$) is shaded in green.} \label{fig: time} 
\end{figure}

We now connect this result to a realistic model for the massive graviton.
The model  setup  is an extension of the Randal-Sundrum (RS) model~\cite{Randall:1999vf, Randall:1999ee, Goldberger:1999uk} with multiple branes~\cite{Kogan:2001qx, Agashe:2016rle, Cai:2021mrw, Cai:2022geu, Lee:2021wau}, where all the SM particles are put on the intermediate  brane $y = r_1 \pi$ with a  tension $T \propto \left (k_{2} - k_{1} \right)$, where $k_{1,2}$ are the curvatures in the two intervals.  To estimate the relic density,  we can ignore the small difference between the two curvatures and take $k \sim M_{pl}$. Hence,  in addition to  a tower of Kaluza-Klein (KK) massive gravitons $G^{(n)}$, the low energy theory  contains a potentially light radion $r$ peaked around the  IR brane $y = r_2 \pi$.  The effective Lagrangian in Eq.~\eqref{Lag0} needs to be extended with the interactions between the massive gravitons and the radion $r$ by expanding the RS metric,
\beq
 \mathcal{L}_{eff} & = &C_{H} \,\sum_i G^{{(n)} \mu \nu} T_{\mu \nu}^{i} + d_V \, r  V_{\mu \nu} V^{\mu \nu}  \nonumber \\ & + & 
C_{r} \left( 2 G^{{(n)} \mu \nu} G^0_{\mu \nu} - G^{(n) \mu}_\mu  G_\nu^{0 \nu} \right) \Box r \nonumber \\ &+& C_{\mathcal{Q}} \mathcal{Q} (G^3, G^2 r, G r^2,  r^3) \,, \label{Lag} 
\eeq
where  $\Box r \equiv  \eta^{\mu \nu} \partial_\mu \partial_\nu r$ and $G^0$ is the massless graviton  (zero mode). The last term in Eq.{\eqref{Lag}} is composed of cubic self-interactions.
The effective couplings for $G^{(n)}$ are derived by the  integration of 5D wave-function overlaps~\cite{Davoudiasl:1999jd, Davoudiasl:2000wi}:
\beq
C_{H} &=& \frac{1}{\Lambda_H} \frac{x_n^2}{4 \sqrt{2} ~ J_2(x_n)}  \,,\\
C_{r}  &=& \frac{1}{M_{pl}} \frac{\sqrt{6} (1- J_0(x_n))}{x_n^2 J_2(x_n)}\,,
\label{coup} 
\eeq
where we applied an approximation in the limit of $e^{- k (r_2 - r_1) \pi} \ll  1$ and defined $\Lambda_{H} = M_{pl} e^{-k (4 r_1 - 3 r_2) \pi}$. The $x_n$ is the root of $J_1(x_n) =0$.  The self interaction strength is characterized by $C_{\mathcal{Q}} \cdot p^2$  and the bulk integration predicts $C_{\mathcal{Q}} \sim 1/\Lambda_{\rm IR} $ with  $\Lambda_{\rm IR} = M_{pl} e^{-k r_2 \pi}$.  The coupling of radion to di-photon originates from the trace anomaly at the loop level:
\beq
d_{\gamma}  &=& \frac{\alpha_{EM}}{8 \pi \sqrt{6}  \Lambda_r}   \left( b_{EM} -  \sum_{i} F_i \right) \,,
\eeq
with the cut off scale  $\Lambda_{r} =  M_{pl} e^{-k (2 r_1 -  r_2) \pi} $ and we used the  electromagnetic  beta function  $b_{EM} =- \frac{11}{3}$, with  $\sum_i F_i \simeq -\frac{1}{9}$  related to the loop functions of $W$  and those heavy fermions  including $t, b, c, s$,$\tau, \mu$~\cite{Giudice:2000av, Csaki:2007ns}. 

The mass  and coupling orders  are of crucial importance for the cosmological stability of the lightest KK graviton as DM.  Note that in this 5D model the coupling  $G^{(1)}$-$G^{0}$-$G^{0}$ is highly suppressed by order of  $M_{pl} e^{k r_2 \pi}$, unlike the bigravity model where this coupling is absent~\cite{Babichev:2016bxi}.  The decay width of  a  massive graviton into two  zero gravitons is  negligible.  Also the hadronic decays are  kinematically forbidden for  $M_G $ in MeV range.   For the lightest  KK graviton, due to the $\mathcal{O}(1)$ self-coupling,  we have to require $ m_r >  M_{G^{(1)}}/2 $ to  shut down the decay of  $G^{(1)} \to 2 r$.   Hence the  relevant decay channels are $G^{(1)} \to e^+ e^-$,  $\nu \bar \nu$ and $\gamma \gamma$ with their decay width given by Eq.~\eqref{tot}.   The other decay patterns will be determined by the mass of radion. Note that from Naive Dimensional Analysis (NDA) a radion mass (or dilaton in the 4D CFT dual) is expected to be around the KK scale~\cite{Bellazzini:2012vz,Chacko:2012sy,Chacko:2013dra}, without the necessity for a fine-tuning. First of all, we calculated the decay width for $G^{(n)} \to G^{0} +r $:
\beq 
\Gamma(G^{(n)} \rightarrow G^{0} r) &=& \frac{C_{r}^2 m_r^4 }{4 \pi M_G} \left(1- \frac{ m_r^2}{M_G^2}\right)\,,
\eeq
with the other relevant decay widths of $G^{(n)} \to 2 G^{(1)}$, $2 r$  and  $r \to 2 G^{(1)}$, $2 \gamma$  given in~\cite{Lee:2013bua, Dillon:2016tqp, Davoudiasl:2001uj}. 

Here we consider two scenarios: in one case, $ M_{G^{(1)}}/2 < m_r \lesssim M_{G^{(2)}}/2$ allows for a Planck-suppressed decay of $G^{(1)} \to G^{0} + r$ and a prompt decay of $G^{(2)} \to 2 r $; in the second, the radion is slightly heavier $ m_r  \gtrsim  2 M_{G^{(1)}} > M_{G^{(2)}}$ so that the radion  and $G^{(n)}$ ($n \geq 3 $) quickly decay into 2 $G^{(1)}$ in less than $10^{-14}$ Sec for $M_{G^{(1)}} \sim 2$ MeV.  As shown in Figure~\ref{fig: time}, for $k r_1 = 11.0$,   we roughly need $k r_2 \gtrsim 14.8 $ to make the  lowest KK graviton to be  stable beyond the Hubble time level. However for the IR brane radion, a much larger value of $k r_2 \gtrsim 16.7$ is required  to barely  ensure the same property. This reflects the fact that the cut off scale $\Lambda_H$ of the KK graviton is of order Planck scale, while the one for the radion is smaller.  For  the lighter radion case,  the dashed red line bends at a certain point, indicating that the  dominant decay becomes $G^{(1)} \to  G^{0} + r$, as the coupling $C_r$ does not decrease with the radius $r_2$ of the IR brane.  A small $m_r$ will  render the graviton $G^{(1)}$ less stable. Thus the second scenario seems to be a more natural option, where  the lightest  two gravitons  are  both long-lived enough to play the role of  DM.

\begin{figure}[tb] 
	\centering 
	\includegraphics[height=5.8 cm, 
		width=7.1 cm]{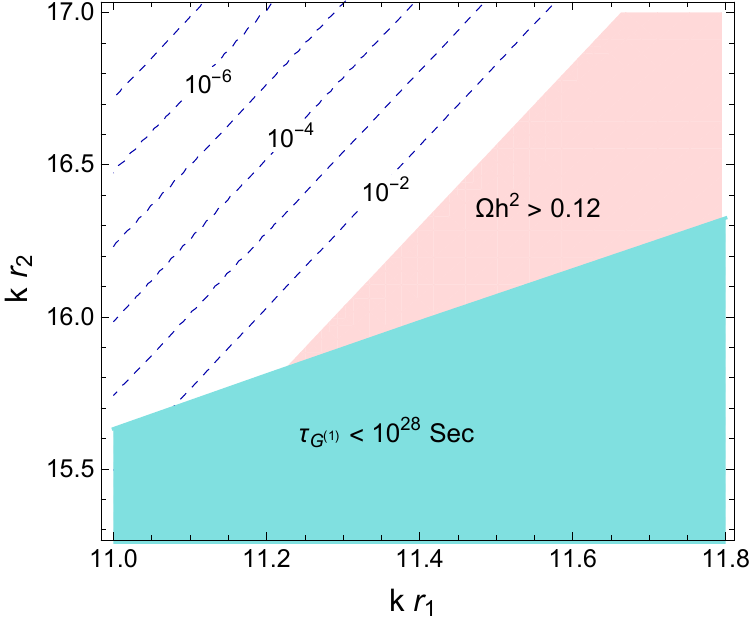}
	\caption{The contour of  relic density in the plane  of $(k r_1, k r_2)$, where the blank region is viable for the graviton as  long-lived  DM. The boundary of  the pink region satisfies $\Omega h^2 = 0.12 $ by summing the direct  and subsequent freeze-in. The light blue region is excluded by the indirect direction.} \label{fig: relic} 
\end{figure}

In the following, we will focus on the heavier radion scenario. The heavier KK gravitons will also undergo a non-thermal IR freeze-in and afterward cascade deposit their density into the stable gravitons.  Since the energy is conserved  during the subsequent freeze-in,  the relic density  approximately is:
\beq
\Omega_{\rm IR } h^2 
& \simeq &  1.75 \times 10^{-62} ~ e^{k (8 r_1 - 3 r_2) \pi} \sum_{n=1}^{10} x_n^2  \,, \label{OmeRS}
\eeq
where we traded $(C_H, M_G)$ in  Eq.~\eqref{OmegaIR} with the two radius parameters in the 5D  realization and summed up all the contribution till  the $n= 10$ KK graviton. Note that the self interactions of the radion receive a very large (finite) 1-loop correction from the KK gravitons, hence the 5D description will no longer be reliable beyond  $10 M_{KK} $ scale.  The prediction of  Figure~\ref{fig: relic} is consistent with the upper limit derived in Eq.~\eqref{bound}.  Since $G^{(1)}$ and $G^{(2)}$   mainly decay into diphoton and neutrinos,  we impose a proper gamma-ray bound  $\tau_{G^{(1)}} \sim 10 \, \tau_{G^{(2)}} > 10^{28}$ sec  for the viable relic density. In Figure~\ref{fig: relic}, the  contours for relic density ($\Omega h^2 =  10^{-2}, 10^{-3}, 10^{-4}, 10^{-5}, 10^{-6}, 10^{-7}$) are parallel to each other and equally spaced. Furthermore, the ratio of the lines is determined by  the exponential factor in Eq.~(\ref{OmeRS}) $  \frac{\partial r2}{\partial r1} = 8/3$.  For $k r_1 \sim 11.2$ and $k r_2 \sim 15.8$,  the  lowest two KK gravitons  can achieve a large fraction of the observed DM relic density.

In summary, in this Letter we presented the effect of a chiral enhancement for the single production of a massive graviton  in the early Universe. This enhancement is active below the EWSB scale and insensitive to the UV physics. Thanks to this novel effect, a generic massive graviton can play the role of a FIMP DM candidate for masses $M_G \lesssim $ a few MeV.  Note that the mass range makes the spin-2 massive graviton an ideal sub-MeV DM particle, which is currently favored by the small scale galaxy structure~\cite{Boylan_Kolchin_2011, Oman:2015xda, Dvorkin:2020xga}. We have applied this new result to an extension of the Randall-Sundrum model with 3 branes.  Depending on the radion mass, only the first two KK gravitons can be long-lived at a cosmological scale, while the higher modes and radion decay very effectively into DM. Note that the  radion was considered as a DM candidate in~\cite{Kolb:2003mm},  but being difficult to account for a heavy relic density. We  have also implicitly assumed that the pair production of massive gravitons is  subdominant due to the gauge invariance of self-interactions.  
This is the first time that a light spin-2 mode, or a class of ``glueball"  interpolated by the conserved energy-momentum tensor of the strong dynamics from  AdS/CFT correspondence~\cite{Maldacena:1997re, Witten:1998qj, Gubser:1998bc},  is shown to be a feasible DM candidate.  
We illustrated that there exists a sizable parameter space where the light  massive gravitons can saturate the DM relic density, while escaping the stringent bounds on decaying DM from indirect detection and the CMB. 

\section*{Acknowledgements}
H.C. and S.L.\ were supported by the National Research Foundation of Korea (NRF) grant funded by the Korea government (MEST) (No. NRF-2021R1A2C1005615). H.C. acknowledges the support of Tsung-Dao Lee Institute, SJTU  where this project is initiated. G.C. and S.L also acknowledge support from the Campus-France STAR project ``Higgs and dark matter connections''.

\bibliography{graviton}
\onecolumngrid
\appendix

\section{Supplementary material: Amplitude Squared for Single Graviton Production}\label{app: amplitude}
For the process  $\bar q (p_1) + q (p_2) \to g (k_1) + G^{\mu \nu}(k_2)$,  there are  a contact term contribution plus the $S$, $T$ and $U$ channels, listed  in order as following:
\beq
\mathcal{M}_1^a &=& - \frac{i}{2} C_H g_s \bar{v}(p_2) \left( \eta_{\mu \rho} \gamma_{\nu} + \eta_{\rho \nu} \gamma_{\mu}\right) T^a u(p_1)  \epsilon^\rho (k_1) \epsilon^{\mu \nu} (k_2) \\
\mathcal{M}_2^a &=&  - i C_H g_s \bar{v}(p_2) \frac{W_{\mu \nu, \sigma \rho} ( -p_1-p_2, k_1)}{(p_1+p_2)^2} \gamma^\sigma T^a u(p_1)   \epsilon^\rho (k_1) \epsilon^{\mu \nu} (k_2) \\
\mathcal{M}_3^a &=& \frac{i}{4} C_H g_s \bar{v}(p_2) \frac{ \gamma_\mu \left(p_1-k_1-p_2  \right)_\nu+ \gamma_{\nu} \left(p_1-k_1-p_2 \right)_\mu  }{(p_1-k_1)^2 -m_q^2} ({\not p_1}- {\not k_1} -m_q) \gamma_\rho T^a u(p_1)  \epsilon^\rho (k_1) \epsilon^{\mu \nu} (k_2) \\
\mathcal{M}_4^a & = & \frac{i}{4} C_H g_s \bar{v}(p_2) \gamma_\rho \left({\not p_1}- {\not k_2} -m_q \right) \frac{ \gamma_\mu \left(p_1-k_2+p_1 \right)_\nu+ \gamma_{\nu} \left(p_1-k_2+p_1 \right)_\mu  }{(p_1-k_2)^2 -m_q^2}  \gamma_\rho T^a u(p_1)  \epsilon^\rho (k_1) \epsilon^{\mu \nu} (k_2)
\eeq
where $T^a$ is the $SU(3)$ gauge generator,  $\epsilon^\rho(k_1)$ and $\epsilon^{\mu \nu}(k_2)$ are the gluon and graviton polarization tensors respectively. When we work in  the $\xi = 1$ gauge,  the symbol $W_{\mu \nu, \sigma \rho} (k_1, k_2) $ including the gauge fixing term is defined as: 
\beq
W_{\mu \nu, \sigma \rho} (k_1, k_2)  =  k_1 \cdot k_2 C_{\mu \nu, \sigma \rho} + D_{\mu \nu, \sigma \rho} (k_1, k_2)+ \xi^{-1} E_{\mu \nu, \sigma \rho} (k_1, k_2)
\eeq
with 
\beq
C_{\mu \nu, \sigma \rho}  &=& \eta_{\mu\sigma}\eta_{\nu\rho} +  \eta_{\mu\rho}\eta_{\nu\sigma}
-\eta_{\mu\nu}\eta_{\sigma \rho}\ ,  \nonumber \\
D_{\mu \nu, \sigma \rho}(k_1, k_2) &=& \eta_{\mu\nu} k_{1\rho}k_{2\sigma} - \biggl[\eta_{\mu\rho} k_{1\nu} k_{2\sigma}
  + \eta_{\mu\sigma} k_{1\rho} k_{2\nu} - \eta_{\sigma \rho} k_{1\mu} k_{2\nu}
  + (\mu\leftrightarrow\nu)\biggr]\ ,  \nonumber \\
E_{\mu \nu, \sigma \rho}(k_1, k_2) &=&   \eta_{\mu\nu}(k_{1\sigma} k_{1\rho}
+ k_{2\sigma} k_{2\rho} +k_{1\sigma}k_{2\rho}) -\biggl[\eta_{\nu\rho}k_{1\mu}k_{1\sigma} +\eta_{\nu\sigma}k_{2\mu}k_{2\rho}+(\mu\leftrightarrow\nu)\biggr] \,.
\eeq
In the limit of $m_q \to 0$,   the amplitude squared can be derived to be:
\beq
|\mathcal{M}_{tot}^a |^2 &=&  |\mathcal{M}_1^a + \mathcal{M}_2^a + \mathcal{M}_3^a + \mathcal{M}_4^a|^2  \nonumber \\
&=& \frac{4 C_H^2 g_s^2}{s t \left(s+t - M_G^2 \right)}  \left( 4 \left(s^2+2 s t+2 t^2\right)  \left(s+t \right) t  - M_G^6 \left(s+4 t \right) \right. \nonumber \\  & & \left.+ 6 M_G^4  \left(s+2 t \right) t- M_G^2
   \left(s^3+6 s^2 t+18 s t^2+16 t^3\right)  \right) 
\eeq
where $s = (p_1+ p_2)^2$ and $t= (p_1-k_1)^2$ are the Mandelstam variables.

We can show how the longitude modes enter into effect.The sum of polarization for a massive graviton is:
\beq
P_{\mu \nu, \alpha \beta} =  \frac{1}{2} \left( P_{\mu \alpha} P_{\nu \beta}  +   P_{\nu \alpha} P_{\mu \beta}   -  \frac{2}{3} P_{\mu \nu} P_{\alpha \beta} \right) 
\eeq
For the tensor $P_{\mu \nu } = \eta_{\mu \nu } -\frac{k_\mu k_\nu}{M_G^2}$,  the  last  piece comes from the longitude polarisation.  We can peel off  this  pure  longitude part $\frac{k_{2,\mu} k_{2, \nu}}{M_G^2}$ so that the respective amplitude is :
 \beq
 \mathcal{M}_{tot}^{a, L} & =& \frac{i C_H g_s \epsilon_\rho (k_1) }{2 M_G^2} \bar v(p_2) \bigg( \left(m_q^2-(p_1-k_1)^2 \right)  \gamma^\rho   + 2 m_q \left( \gamma^\rho  \not k_2  -   \not k_2  \gamma^\rho \right)  \nonumber \\  &    - &   2   \not k_2  \left( p_2^\rho - k_1^\rho \right) - 2   \not k_1 \left(k_2^\rho - k_1^\rho M_G^2/s \right)  -   \not k_2 \not k_1 \gamma^\rho   \bigg) T^a u(p_1) 
 \eeq
where we keep the terms proportional to $k_1^\rho$.  For $m_q = 0$, because  the longitude mode will not contribute to the cross section due to helicity conservation,  we will get $ |M_{tot}^{a, L}|^2 = 0$. However after the quark obtains mass, the amplitude squared with the color factor ${\rm Tr} \, T^a T^a = 4$ counted is:
 \beq
|\mathcal{M}_{tot}^{a, L}|^2 &=&  \frac{128 C_H^2 g_s^2 m_q^2 \left(m_q^2-t\right)
   \left(s+t -m_q^2 \right)}{3 M_G^4}  \nonumber \\ &+& \mathcal{O} (\frac{1}{M_G^2})
 \eeq
 As we can see there is an enhancement factor  $m_q^4/M_G^4$ in the amplitude squared provided the  mass of  graviton is in the KeV-MeV order. This type of enhancement also applies to the process of $q \bar{q} \to h G^{(1)}$. However for the latter, its amplitude squared is proportional to $C_H^2 \frac{ m_q^4 s}{v^2 M_G^2}$,  thus contributes less  to the freeze-in relic density.

\end{document}